\documentclass[conference]{IEEEtran}
\IEEEoverridecommandlockouts

\usepackage{cite}
\usepackage{amsmath,amssymb,amsfonts}
\usepackage{graphicx}
\usepackage{textcomp}
\usepackage{xcolor,pifont}
\usepackage{algorithm, algorithmic}

\usepackage{mathtools}
\usepackage{multirow}
\usepackage{subfigure}
\usepackage{algorithm, algorithmic}
\usepackage{bm}
\usepackage{comment}
\algsetup{linenosize=\small}

\usepackage{tikz}

\usepackage{mathabx}

\newcommand*\colourcheck[1]{%
  \expandafter\newcommand\csname #1check\endcsname{\textcolor{#1}{\ding{52}}}%
}
\colourcheck{green}

\newcommand*\colourcross[1]{%
  \expandafter\newcommand\csname #1cross\endcsname{\textcolor{#1}{\ding{56}}}%
}
\colourcross{red}

\def\BibTeX{{\rm B\kern-.05em{\sc i\kern-.025em b}\kern-.08em
    T\kern-.1667em\lower.7ex\hbox{E}\kern-.125emX}}

\newcommand{\pass}[1]{\textcolor{cyan}{#1}}

\begin{document}

\title{Leveraging FPGAs for Homomorphic Matrix-Vector Multiplication in Oblivious Message Retrieval}

\author{\IEEEauthorblockN{Grant Bosworth\IEEEauthorrefmark{1},
Keewoo Lee\IEEEauthorrefmark{2}, and
Sunwoong Kim\IEEEauthorrefmark{1}}
$^*$Rochester Institute of Technology, Rochester, NY, USA, $^\dagger$Ethereum Foundation \\
Email: \{grb1300, sskeme\}@rit.edu, keewoo.lee@ethereum.org}

\maketitle

\begin{abstract}
While end-to-end encryption protects the content of messages, it does not secure metadata, which exposes sender and receiver information through traffic analysis. A plausible approach to protecting this metadata is to have senders post encrypted messages on a public bulletin board and receivers scan it for relevant messages. Oblivious message retrieval (OMR) leverages homomorphic encryption (HE) to improve user experience in this solution by delegating the scan to a resource-rich server while preserving privacy. A key process in OMR is the homomorphic detection of pertinent messages for the receiver from the bulletin board. It relies on a specialized matrix-vector multiplication algorithm, which involves extensive multiplications between ciphertext vectors and plaintext matrices, as well as homomorphic rotations. The computationally intensive nature of this process limits the practicality of OMR. To address this challenge, this paper proposes a hardware architecture to accelerate the matrix-vector multiplication algorithm. The building homomorphic operators in this algorithm are implemented using high-level synthesis, with design parameters for different parallelism levels. These operators are then deployed on a field-programmable gate array platform using an efficient design space exploration strategy to accelerate homomorphic matrix-vector multiplication. Compared to a software implementation, the proposed hardware accelerator achieves a 13.86$\times$ speedup.
\end{abstract}


\section{Introduction}
%
End-to-end encryption (E2EE) ensures only the sender and receiver can access a message, blocking even the service provider. 
However, E2EE does not effectively protect metadata, which can reveal \emph{who communicated with whom} \cite{liu2022oblivious}. 
Metadata often enables detailed tracking of online activity, sometimes making access to message content unnecessary.
A key aspect of metadata protection is receiver privacy, aiming to conceal the receiver's identity.

A plausible approach to achieving receiver privacy is to have senders post encrypted messages on a public bulletin board, allowing receivers to scan it for messages intended for them. However, this approach imposes a significant burden on receivers. To improve usability, several works have proposed delegating this costly linear scan to a server in a privacy-preserving way. Yet, these solutions either provide only a weak form of privacy \cite{beck2021fuzzy} or rely on strong environmental assumptions, such as the availability of trusted execution environments or multiple communicating-but-non-colluding servers \cite{madathil2022private, jakkamsetti2025scalable}.

To achieve full receiver privacy under minimal assumptions, oblivious message retrieval (OMR) was introduced by Liu and Tromer \cite{liu2022oblivious}, by incorporating homomorphic encryption (HE), which is a cryptographic technique that allows computations on encrypted data directly \cite{cheon2021introduction}.
HE enables offloading the task of detecting payloads pertinent to the receiver, among those from multiple senders, from the receiver to a resource-rich third-party server.
Since the introduction of the original OMR, several variants have been proposed, particularly focused on improving the efficiency of server-side operations \cite{liu2024perfomr, lee2024sophomr}.

Despite its potential in various applications, such as privacy-preserving cryptocurrencies, OMR faces several practical challenges.
One major issue is the slow processing speed introduced by the use of an HE scheme.
To address this, we propose a field-programmable gate array (FPGA)-based accelerator for homomorphic matrix-vector multiplication, which is a key performance bottleneck in OMR.
Our accelerator achieves a performance improvement of 13.86$\times$ over a CPU implementation.
To the best of the authors' knowledge, this is the first work on custom hardware accelerators for OMR.

\section{Background}
%
\subsection{BFV Homomorphic Encryption Scheme}
%
The Brakerski–Fan–Vercauteren (BFV) scheme is one of the most widely used HE schemes, supporting integer arithmetic modulo a prime number \cite{brakerski2012fully, fan2012somewhat}.
It consists of the following core stages: key generation, encoding, encryption, evaluation, decryption, and decoding.
Key generation creates secret, public, and evaluation keys, based on HE parameters.
Encoding takes multiple integer messages and packs them into a vector, which is then transformed into a plaintext polynomial \texttt{pt} $\in \mathcal{R}_t$ = $\mathbb{Z}_t[X]/(X^n+1)$, where $t$ is the plaintext modulus and $n$ is the ring dimension.
Encryption uses the public key and the plaintext polynomial to generate a ciphertext \texttt{ct} $\in \mathcal{R}_Q \times \mathcal{R}_Q$ (i.e., two ciphertext polynomials), where $\mathcal{R}_Q$ = $\mathbb{Z}_Q[X]/(X^n+1)$ and $Q$ is the ciphertext modulus.
Since $Q$ is often hundreds or thousands of bits long, large-number arithmetic is required, which is computationally inefficient in standard computing environments.
To address this, the residue number system (RNS) is applied to the BFV scheme \cite{bajard2016full}. 
In RNS-based BFV, each ciphertext polynomial is decomposed into several smaller polynomials (known as limbs), each modulo a smaller prime $q_i$, such that $Q = \Pi_i q_i$.
These smaller polynomials can be processed independently, while still producing the same result as using the original $Q$.

Evaluation uses homomorphic operations to carry out a specific privacy-preserving application.
The following are the BFV homomorphic operations, which are used in this work:
\begin{itemize}
    \item \textbf{Ciphertext-ciphertext addition} (\texttt{CCadd}) performs element-wise addition of the corresponding polynomials in two ciphertexts. Given ciphertexts \texttt{ct}$_1$ = (\texttt{ct}$_1^{(1)}$, \texttt{ct}$_1^{(2)}$) and \texttt{ct}$_2$ = (\texttt{ct}$_2^{(1)}$, \texttt{ct}$_2^{(2)}$), \texttt{CCadd}(\texttt{ct}$_1$, \texttt{ct}$_2$) is ((\texttt{ct}$_1^{(1)}$ + \texttt{ct}$_2^{(1)}$)$\mod q$, (\texttt{ct}$_1^{(2)}$ + \texttt{ct}$_2^{(2)}$)$\mod q$) = \texttt{ct}$_3$.

    \item \textbf{Plaintext-ciphertext multiplication} (\texttt{PCmul}) multiplies a plaintext polynomial by a ciphertext, producing a new ciphertext. Given a ciphertext \texttt{ct} = (\texttt{ct}$^{(1)}$, \texttt{ct}$^{(2)}$) and a plaintext \texttt{pt}, \texttt{PCmul}(\texttt{pt}, \texttt{ct}) is defined as ((\texttt{ct}$^{(1)}$ $\cdot$ \texttt{pt})$\mod q$, (\texttt{ct}$^{(2)}$ $\cdot$ \texttt{pt})$\mod q$)) = \texttt{ct}$'$.

    \item \textbf{Rotation operation} (\texttt{Rot}) shifts the slots of an encrypted vector. Let \texttt{ct} be a ciphertext encrypting a vector of integers $\bf{v}$ = \{$v_0, v_1, ..., v_{u-1}$\}. \texttt{Rot}$_a$(\texttt{ct}) denotes a rotation of the ciphertext by \texttt{a} slots, resulting in a new ciphertext that encrypts the rotated vector \{$v_a, v_{a+1}, ..., v_{a-1}$\}.
    \texttt{Rot} consists of two main steps: \texttt{ApplyGalois} and \texttt{KeySwitch}. \texttt{ApplyGalois} applies a Galois automorphism to the ciphertext, which rearranges its coefficients.
    \texttt{KeySwitch} uses rotation keys (also known as Galois keys), a type of evaluation keys, to transform the resulting ciphertext back into a form that can be decrypted using the original secret key.
    Compared to \texttt{CCadd} and \texttt{PCmul}, \texttt{Rot} is computationally more expensive. This is mainly due to the \texttt{KeySwitch} step, which involves modular multiplications with large rotation keys and number theoretic transform (NTT). 
    For more details, refer to \cite{ozcan2023homomorphic}.
\end{itemize}
Decryption uses the secret key to recover the corresponding plaintext polynomial from the resulting ciphertext.
Decoding then transforms this polynomial into a plaintext vector, from which the resulting messages are extracted.

\subsection{Oblivious Message Retrieval}
\begin{figure}[t]
    \centering
    \includegraphics[scale=0.38]{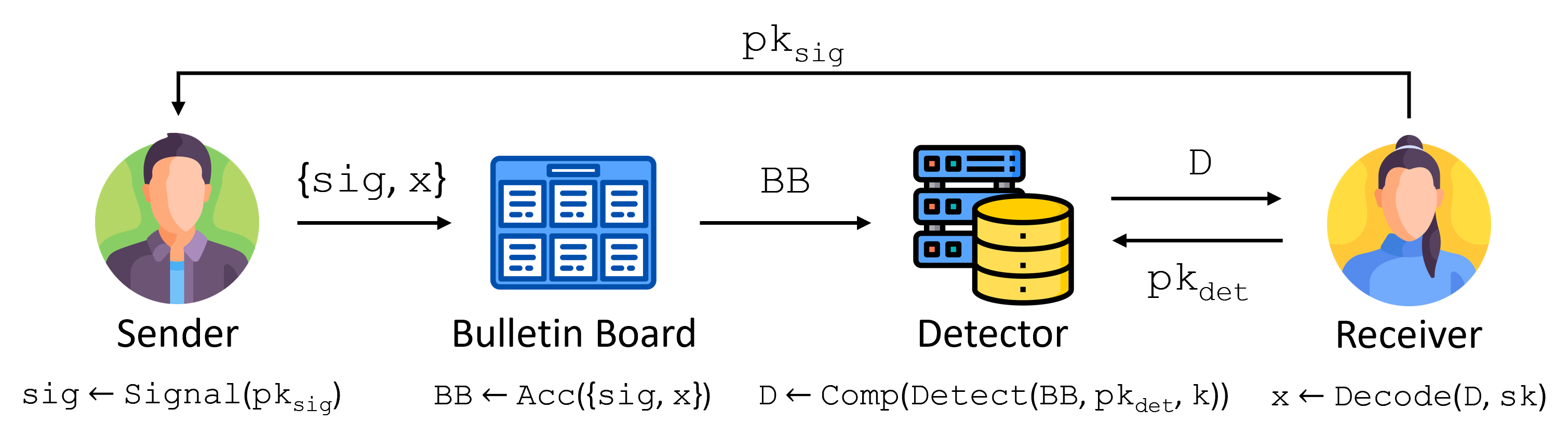}
    \caption{Overall flow of SophOMR based on a PS technique.}
    \label{fig:OMR_arch}
    \vspace{-1em}
\end{figure}

%
Private signaling (PS) introduced in \cite{lee2024sophomr} is a method for transmitting payloads intended for a specific receiver, while concealing the intent of a sender from others.
Fig. \ref{fig:OMR_arch} shows a secure communication process based on PS.
Upon receiving a public signal key \texttt{pk$_{\texttt{sig}}$} from the receiver, the sender generates a signal \texttt{sig} as a clue for the payload \texttt{x}.
The sender then posts the \texttt{sig}-\texttt{x} pair on a public bulletin board \texttt{BB}, which accumulates $N$ signal-payload pairs from various senders.
The signal is cryptographically constructed so that only the party possessing the secret key (i.e., the intended receiver) can recognize that it is meant for them, while revealing no information otherwise.
Although the receiver could download the entire \texttt{BB} and decrypt all pairs to identify the pertinent payloads, this approach is computationally intensive and imposes significant communication overhead on the receiver.

To address these issues, OMR delegates the scanning task to a resource-rich third-party server known as the detector \cite{liu2022oblivious}. 
OMR employs an HE scheme to detect payloads pertinent to the receiver without exposing any meaningful information to the detector.
In existing OMR schemes \cite{liu2022oblivious, liu2024perfomr, lee2024sophomr}, the receiver typically provides the detector with a detection key \texttt{pk$_{\texttt{det}}$}, which contains the PS secret keys encrypted under the HE scheme, the public keys for the HE scheme, and an upper bound $k$ indicating the maximum number of pertinent payloads the receiver expects to retrieve.

In SophOMR \cite{lee2024sophomr}, which is the state-of-the-art OMR scheme, the detector performs two main tasks: homomorphic detection (\texttt{Detect}) and homomorphic compression (\texttt{Comp}).
\texttt{Detect} homomorphically computes the indices of pertinent payloads from the bulletin board using the detection key.
The results are ciphertexts of a sparse binary vector and serve to mask the pertinent payloads while zeroing out the non-pertinent ones.
Although the resulting data is still as large as the original bulletin board, \texttt{Comp} reduces the size.
It compresses the encrypted sparse vector into a compact digest \texttt{D}, which is significantly smaller than the bulletin board and ideally proportional to $k$.
Since the signal is cryptographically designed to hide the receiver and the operations involved in \texttt{Detect} and \texttt{Comp} are performed in the HE domain, the detector is unable to determine which payloads are relevant to the receiver.
When the digest is sent to the receiver, they decode it using their secret key.

\subsection{Matrix-Vector Multiplication Algorithm in SophOMR}
%
One of the most critical operations in SophOMR is the homomorphic matrix-vector multiplication (\texttt{MatMul}), which multiplies a plaintext matrix $\bf{M}$ of size $N \times k$ with a ciphertext $\bf{v}$ encrypting a vector of length $k$, as follows:
\begin{equation}
\label{eq:matmul}
    \begin{split}
    \textbf{Mv} = \Sigma_{i=0}^{k-1} \texttt{diag}_i(\textbf{M}) \odot \texttt{Rot}^i(\textbf{v}),
    \end{split}
\end{equation}
where $\texttt{diag}_i(\textbf{M})[j]$ = $\textbf{M}[j \text{ mod } N][(i+j) \text{ mod } k]$ and $\odot$ represents multiplication between a plaintext and a ciphertext.
To optimize this operation, SophOMR applies the baby-step gaint-step style optimization \cite{halevi2018faster} to \texttt{MatMul}, as follows:
\begin{equation}
\label{eq:bsgs_matmul}
    \begin{split}
    \textbf{Mv} = \Sigma_{g=0}^{\tilde{g}-1} \texttt{Rot}^{g \tilde{b}} (\Sigma_{b=0}^{\tilde{b}-1} \textbf{m}_{g \tilde{b} + b} \odot \texttt{Rot}^b(\textbf{v})),
    \end{split}
\end{equation}
where $k$ = $\tilde{g} \cdot \tilde{b}$ and $\textbf{m}_{g\tilde{b}+b}$ denotes \texttt{Rot}$^{-g\tilde{b}}$(\texttt{diag}$_{g\tilde{b}+b}(\bf{M})$).
This optimization reduces the number of \texttt{Rot} operations from $k = \tilde{g} \cdot \tilde{b}$ to $\tilde{g} + \tilde{b}$, while requiring only two rotation keys.
The optimized \texttt{MatMul} algorithm is presented in Algorithm \ref{alg:matmul}.
\begin{algorithm}[t]
    \caption{\texttt{MatMul} \cite{lee2024sophomr}}
    \label{alg:matmul}
    \begin{algorithmic}[1]
    \renewcommand{\algorithmicrequire}{\textbf{Input:} ($\bf{m}$$_j$)$_{j=0}^{k-1}$, \texttt{ct}$_{\text{in}}$} 
    \REQUIRE
    \renewcommand{\algorithmicrequire}{\textbf{Output:} \texttt{ct}$_{\text{out}}$}
    \REQUIRE
    
    \STATE $\texttt{ct}_0 \leftarrow \texttt{ct}_{\text{in}}$
    \FOR{($b=1$; $b<\tilde{b}$; $b=b+1$)}
        \STATE $\texttt{ct}_b$ $\leftarrow$ \texttt{Rot}$^1$(\texttt{ct}$_{b-1}$)
    \ENDFOR
    
    \FOR{($g=\tilde{g}-1$; $g \geq 0$; $g=g-1$)}
        \STATE \texttt{ct}$_{\text{sum}}$ $\leftarrow$ $\Sigma_{b=0}^{\tilde{b}-1}$ \texttt{PCmul}($\bf{m}$$_{ g\tilde{b}+b}, $ \texttt{ct}$_b$)
        \IF{$g = \tilde{g}-1$}
            \STATE \texttt{ct}$_{\text{out}}$ $\leftarrow$ \texttt{ct}$_{\text{sum}}$
        \ELSE
            \STATE \texttt{ct}$_{\text{out}}$ $\leftarrow$ \texttt{CCadd}(\texttt{Rot}$^{\tilde{b}}$(\texttt{ct}$_{\text{out}}$), \texttt{ct}$_{\text{sum}})$
        \ENDIF
    \ENDFOR
    
    \STATE \textbf{return} \texttt{ct}$_{\text{out}}$ \\
    \end{algorithmic} 
\end{algorithm}
%

\section{Proposed Work}
%

\subsection{Motivation}
\label{Motivation}
\begin{figure}[t]
	\centering
	\subfigure[]{
	    \includegraphics[scale=0.195]{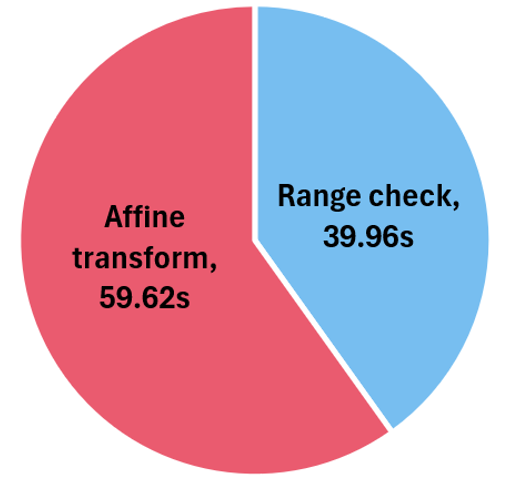}	    
	    \label{fig:breakdown_a}
	}
	\subfigure[]{
	    \includegraphics[scale=0.195]{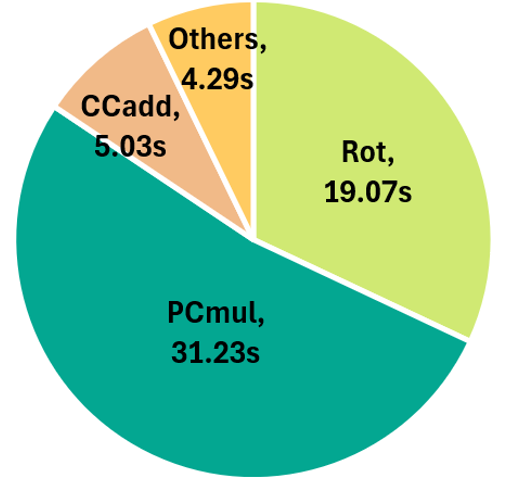}	    
	    \label{fig:breakdown_b}
	}
	\subfigure[]{
	    \includegraphics[scale=0.195]{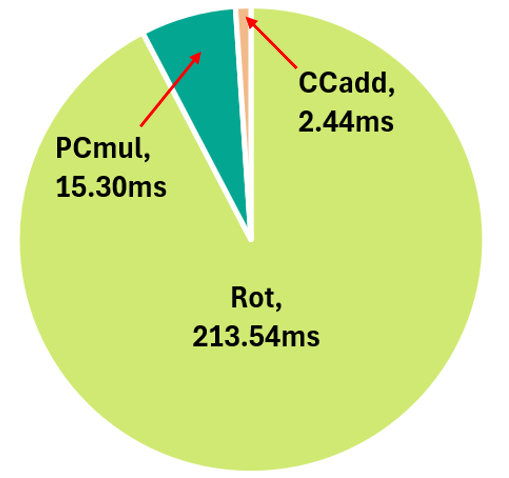}	    
	    \label{fig:breakdown_d}
	}
	\caption{CPU time breakdown of \texttt{Detect} when $k$ = 50 and a single thread is used. (a) Time distribution within \texttt{Detect} for $N$ = $2^{16}$. (b) Time distribution within \texttt{MatMul} operations. (c) Execution time of individual HE operations.}
	\label{fig:breakdown}
	\vspace{-1.5em}
\end{figure}

%
This paper focuses on accelerating the detector operations of SophOMR.
Specifically, we target the \texttt{MatMul} operation in the \texttt{Detect} phase, which offers high potential for parallelism.
The \texttt{Detect} phase is divided into two components: an affine transform and a range check.
%
%
While the range check involves more consecutive homomorphic multiplications, a key factor in the performance of HE schemes, the affine transform often exhibits longer execution times.
Fig. \ref{fig:breakdown_a} shows the time distribution within the \texttt{Detect} phase with a bulletin board size ($N$) of $2^{16}$, evaluated on a workstation equipped with an Intel Xeon W-2295 processor and 128 GB of RAM.
The affine transform takes approximately 1.5$\times$ longer to execute compared to the range check.
%
%

Nearly all the time spent in the affine transform is consumed by \texttt{MatMul} operations. 
Fig. \ref{fig:breakdown_b} shows a breakdown of the execution time for the \texttt{MatMul} operations. 
More than 50\% of the time is spent on \texttt{PCmul} operations, followed by \texttt{Rot} operations, which account for more than 30\%. 
However, when looking at the execution time of a single operation, \texttt{Rot} takes significantly longer than \texttt{PCmul}, as shown in Fig. \ref{fig:breakdown_d}.
This implies that accelerating \texttt{MatMul} requires (1) optimizing the \texttt{Rot} operation itself and (2) parallelizing the large number of \texttt{PCmul} operations.
To address them, we use multi-level design parameters along with a design space exploration (DSE) technique to find their optimal values.

\subsection{Overall Design Process}
\begin{figure}[t]
    \centering
    \includegraphics[scale=0.4]{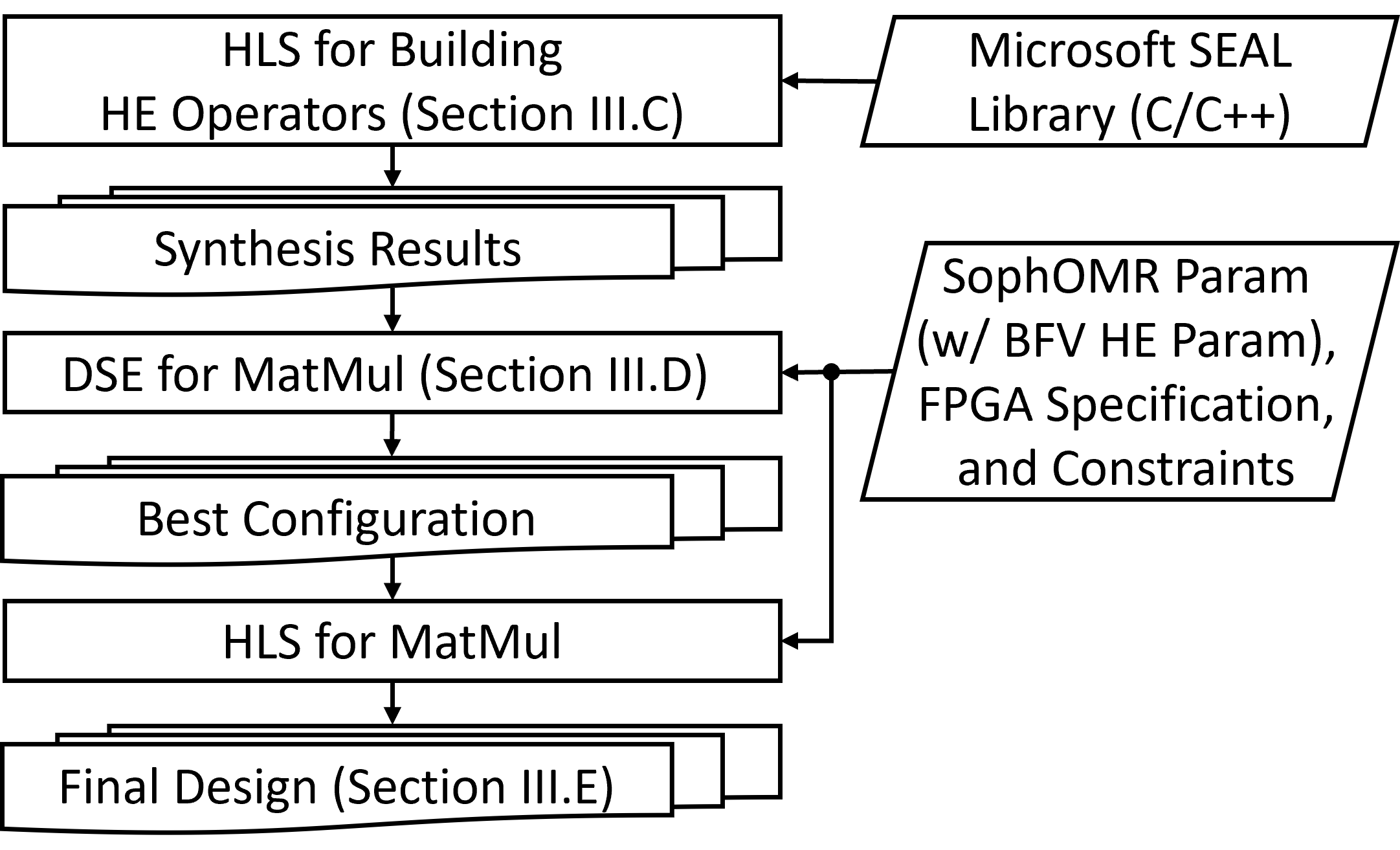}
    \caption{Overall design process to design our \texttt{MatMul} accelerator.}
    \label{fig:overall_flow}
    \vspace{-1em}
\end{figure}

%
Our DSE-based design process is shown in Fig. \ref{fig:overall_flow}.
We first use high-level synthesis (HLS) to get the synthesis results, specifically the latency and hardware resource usage, of each building homomorphic operator used in \texttt{MatMul}. 
To this end, we extracted the relevant functions from the Microsoft SEAL open-source library \cite{sealcrypto} and refactored them for HLS compatibility.
Note that although the SophOMR open-source code \cite{sophomr_github} is based on the OpenFHE library \cite{OpenFHE}, this paper opts for SEAL due to its clearer function hierarchies, which simplify the generation of HLS code.
Each homomorphic operator is implemented in multiple versions corresponding to different parameter configurations. 
The parameter types used in this work are summarized in Table \ref{tab:parallelism_type} and described in detail in the following subsections.
\begin{table}[t]
	\centering
	\caption{Design Parameters for \texttt{MatMul} Accelerators}
	\label{tab:parallelism_type}
	\begin{tabular}{c | c | c}
		\hline
        Parallelism type & Related homomorphic operator & Parameter \\
        \hline
        Coefficient level & \texttt{CCadd}, \texttt{PCmul}, \texttt{Rot} & $PC$ \\
        NTT BU level & \texttt{Rot} & $PB$ \\
        Instance level & \texttt{PCmul} & $PI$ \\
		\hline
	\end{tabular}
	\vspace{-1em}
\end{table}

Using the synthesis results of the building homomorphic operators together with the cost models presented in Section \ref{Cost Models}, DSE is performed for \texttt{MatMul}.
Compared to general approaches that rely on coarse-grained modeling and extrapolation to estimate latency and hardware resource usage across design variants, our synthesis-based methodology enables more accurate DSE.
During the DSE process, SophOMR parameters, including BFV HE parameters, FPGA specifications, and design constraints are taken into account.
Once the optimal feasible design parameters are determined, HLS is reused to generate the final \texttt{MatMul} accelerator.

\subsection{Implementation of Building Homomorphic Operators}
\label{Implementation of Building Homomorphic Operators}
%
For the \texttt{CCadd} and \texttt{PCmul} modules, the HLS \texttt{PIPELINE} pragma with a target initiation interval of 1 is used to maximize throughput. 
In addition, $PC$ coefficients are processed in parallel to further improve performance.
%
The design parameter $PC$ is also applied to the \texttt{Rot} module, but an additional design parameter is introduced for the \texttt{KeySwitch} module.
%
%
Within the \texttt{KeySwitch} module, the most time-consuming operations are the NTT and the modular multiplications with rotation keys, which are performed in series. 
%
%
As in \cite{zhu2023fxhenn}, our \texttt{KeySwitch} module incorporates limb-based pipelining, allowing the NTT and modular multiplications to be performed in parallel. 
Since the execution time of the NTT is generally longer than that of the modular multiplication, the overall performance of the \texttt{KeySwitch} module is primarily determined by the NTT latency.
The performance of the NTT is directly influenced by the number of butterfly units (BUs) operating concurrently \cite{kim2020hardware}. 
Therefore, the number of BUs is defined as a design parameter $PB$.
For NTT implementation, the HLS-based open-source autoNTT library \cite{kumarathunga2025autontt} is used, specifically the iterative architecture + Barrett reduction version.

\subsection{Cost Models}
\label{Cost Models}
%
Although HLS facilitates the search for optimal design parameters, the design space grows exponentially with the number of tunable parameters. 
To address this, we define cost models to quickly estimate latency and hardware resource usage in the \texttt{MatMul} accelerator, avoiding the need to run HLS for every design configuration.
Note that, following the approach in \cite{zhu2023fxhenn}, only digital signal processing (DSP) slices, block RAM (BRAM), and UltraRAM (URAM) are considered for hardware resource cost models, as FPGAs typically have abundant lookup tables (LUTs) and flip-flops (FFs).

\subsubsection{Latency}
The latency model incorporates the latency synthesis results for the building homomorphic operators and the design parameters listed in Table \ref{tab:parallelism_type}.
We begin by defining the latency for each iteration for $g$ in Algorithm \ref{alg:matmul}, without considering any design parameters.
The \texttt{PCmul} operations and the \texttt{CCadd} operations for accumulation in line 6 are executed in parallel and pipelined manner.
Since the latency of \texttt{PCmul} ($L_M$) is typically longer than the latency of \texttt{CCadd} ($L_A$), the latency for line 6 is set to $\tilde{b} \cdot L_{M} + L_{A}$.
The \texttt{Rot} operation with a rotation amount of $\tilde{b}$ (line 10) is executed in parallel with the \texttt{PCmul} and \texttt{CCadd} operations (when $g \neq \tilde{g}-1$) since there are no data dependencies between them.
Letting the latency of \texttt{Rot} be $L_R$, the latency for each iteration ($IL$) is defined as follows:
\begin{equation}
\label{eq:latency_orig_inner}
    \begin{split}
    IL = \text{max}\{\tilde{b} \cdot L_{M} + L_{A}, L_{R}\} + L_{A}.
    \end{split}
\end{equation}
%
%
Using (\ref{eq:latency_orig_inner}), the total latency of a single \texttt{MatMul} operation is estimated as follows:
\begin{equation}
\label{eq:latency_orig}
    \begin{split}
    TL = (\tilde{b}-1) \cdot L_R + \tilde{g} \cdot IL.
    \end{split}
\end{equation}

By applying the design parameters to (\ref{eq:latency_orig_inner}) and (\ref{eq:latency_orig}), we obtain (\ref{eq:latency_modified_inner}) and (\ref{eq:latency_modified}), respectively.
\begin{equation}
\label{eq:latency_modified_inner}
    \begin{split}
    IL' = \text{max}\{\frac{\tilde{b}}{PI} \cdot L_{M}' + L_{A}', L_{R}'\} + L_{A}',
    \end{split}
\end{equation}
\begin{equation}
\label{eq:latency_modified}
    \begin{split}
    TL' = (\tilde{b}-1) \cdot L_R' + \tilde{g} \cdot IL',
    \end{split}
\end{equation}
where $PI$ denotes the number of \texttt{PCmul} cores operating in parallel, which is detailed in Section \ref{Hardware Architecture}.
$L_M'$ and $L_A'$ are approximately $\frac{L_M}{PC}$ and $\frac{L_A}{PC}$, respectively.
For $L_R'$, $PB$ is applied to NTT and inverse NTT (INTT), while $PC$ is applied to other sub-modules of \texttt{KeySwitch}.
The values of $L_M'$, $L_A'$, and $L_R'$ are obtained directly from the synthesis results.


\subsubsection{DSP Usage}
Similar to latency, the total usage of DSP slices is estimated based on the synthesis results of the building homomorphic operators.
Let $d_{op}$ represent the number of DSP slices used by a homomorphic operator $op \in$ \{\texttt{CCadd}, \texttt{PCmul}, \texttt{Rot}\}.
Then, the total DSP usage $D$ is calculated as follows:
\begin{equation}
\label{eq:lutffdsp}
    \begin{split}
    D = (PI+1) \cdot d_{\texttt{CCadd}} + PI \cdot d_{\texttt{PCmul}} + d_{\texttt{Rot}}.
    \end{split}
\end{equation}

\subsubsection{BRAM and URAM Usage}
In our \texttt{MatMul} accelerator, only the \texttt{Rot} module utilizes BRAMs, specifically within the NTT and INTT modules.
URAMs, on the other hand, serve as internal buffers within the \texttt{Rot} module, including rotation key buffers, and are also used for matrix buffers.
In addition, URAMs are utilized as data transfer buffers to facilitate communication with off-chip memory.
The total URAM usage is obtained by adding all individual values.

\subsection{Hardware Architecture}
\label{Hardware Architecture}
\begin{figure}[t]
    \centering
    \includegraphics[scale=0.35]{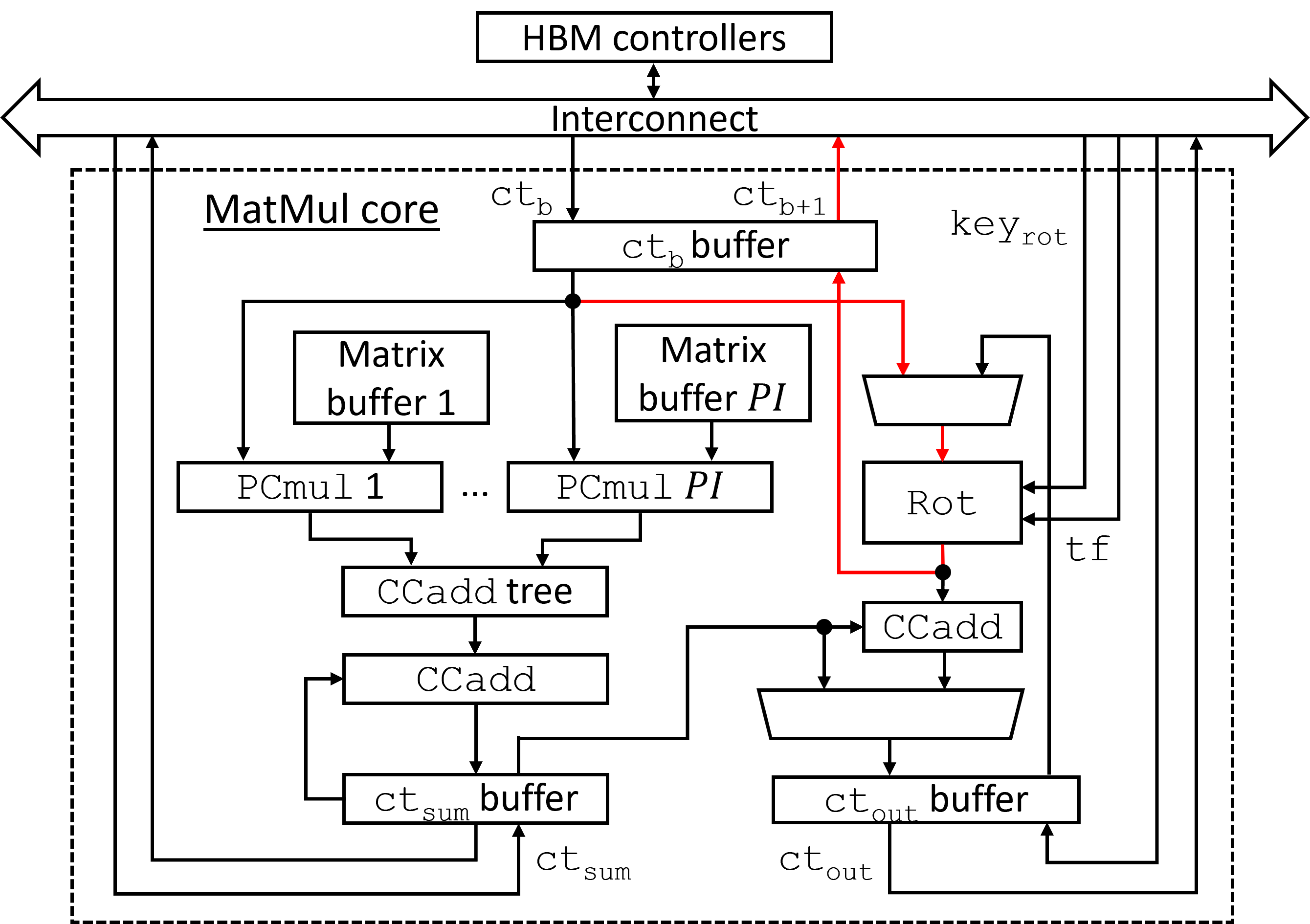}
    \caption{Simplified \texttt{MatMul} hardware architecture. The red arrows represent the data flow corresponding to lines 2-4 of Algorithm \ref{alg:matmul}.}
    \label{fig:matmul_arch}
    \vspace{-1em}
\end{figure}

%
%
Fig. \ref{fig:matmul_arch} shows a simplified block diagram of the proposed \texttt{MatMul} hardware architecture.
A single \texttt{Rot} core is instantiated and shared by lines 3 and 10 of Algorithm \ref{alg:matmul}.
Within the \texttt{Rot} module, buffers are allocated to store rotation keys.
Under the SophOMR parameters (see Table \ref{tab:he_parameter}), a single rotation key occupies 55MB.
Since \texttt{MatMul} uses two rotation keys (for shift amounts 1 and $\tilde{b}$), the total storage of 110MB exceeds the available on-chip memory.
Consequently, only a portion of the keys is kept in URAMs and continuously updated with data fetched from off-chip memory.
%
%
Similarly, the twiddle factors (TFs) used in the (I)NTT are updated from off-chip memory.

The $\tilde{g} \cdot \tilde{b}$ \texttt{PCmul} operations in \texttt{MatMul} are independent of each other.
Therefore, the proposed hardware architecture deploys $PI$ \texttt{PCmul} cores running in parallel.
Accordingly, ($PI-1$) \texttt{CCadd} cores are used to sum the products from the $PI$ \texttt{PCmul} cores, and two additional \texttt{CCadd} cores are used for the accumulation and final ciphertext computation, respectively.
Each \texttt{PCmul} core is accompanied by a dedicated matrix buffer.
Each $\bf{m}$$_{j}$ in Algorithm \ref{alg:matmul} is a single plaintext that remains constant.
When using the SophOMR parameters (in Table \ref{tab:he_parameter}), its size is approximately 1,280Kb.
%
%
The \texttt{MatMul} algorithm involves $k$ matrices, which are precomputed and distributed across the $PI$ matrix buffers.

As described in Section \ref{Cost Models}, data transfer buffers implemented using URAMs (\texttt{ct}$_\texttt{b}$, \texttt{ct}$_\texttt{sum}$, and \texttt{ct}$_\texttt{out}$ buffers) are deployed to temporarily store ciphertexts during transfers to/from off-chip memory.
Due to the large size of a ciphertext, which includes multiple limbs, only a single limb is stored in each data transfer buffer at a time.
To hide off-chip memory latency, a double buffering scheme is utilized.

\section{Evaluation}
%
%
\subsection{Experimental Setup}
%
The SophOMR parameters are listed in Table \ref{tab:he_parameter}.
\begin{table}[t]
	\centering
	\caption{SophOMR Parameters}
	\label{tab:he_parameter}
	\begin{tabular}{c | c | c | c | c | c | c | c}
		\hline
        $N$ & $t$ & $\lceil \log Q \rceil$ & $\lceil \log PQ \rceil$ & $\lceil \log q_i \rceil$ & $k$ & $\tilde{g}$ & $\tilde{b}$ \\
        \hline
        $2^{16}$ & 786,433 & 1,140 & 1,740 & 60 & 50 & 23 & 46 \\
		\hline
	\end{tabular}
	\vspace{-1.3em}
\end{table}
Here, $N$ represents the total number of payloads posted on the bulletin board, which is equal to the BFV ring dimension $n$.
The plaintext and ciphertext moduli in the BFV scheme are denoted by $t$ and $Q$, respectively. 
$P$ is a special modulus used temporarily during the \texttt{KeySwitch} process.
%
%
The target FPGA platform is an AMD Alveo U55C Accelerator Card, which includes 1,304K LUTs, 2,607K FFs, 2,016 36Kb BRAMs, 960 288Kb URAMs, 9,024 DSP slices, and 16GB HBM2.
%
%
The AMD Vitis HLS (v2024.1) is used as a development tool.

\subsection{Hardware Implementation Results}
%
Table \ref{tab:best_configuration} shows the top four parameter configurations for the \texttt{MatMul} accelerator, ranked with latency as the highest priority and hardware resource usage as the second, under the constraint of available hardware resources.
\begin{table}[t]
	\centering
	\caption{Top Four Configurations from DSE for \texttt{MatMul} Accelerators}
	\label{tab:best_configuration}
	\begin{tabular}{c | c | c | c | c | c}
		\hline
        \multirow{2}{*}{Order} & \texttt{CCadd} & \multicolumn{2}{c|}{\texttt{PCmul}} & \multicolumn{2}{c}{\texttt{Rot}} \\
        \cline{2-6}
        & $PC$ & $PC$ & $PI$ & $PC$ & $PB$ \\
        \hline
        1 (best) & 16 & 16 & 2 & 16 & 64 \\
        2 & 16 & 8 & 4 & 16 & 64 \\
        3 & 16 & 4 & 8 & 16 & 64 \\
        4 & 16 & 2 & 16 & 16 & 64 \\
		\hline
	\end{tabular}
	\vspace{-2em}
\end{table}
These configurations show almost identical \texttt{MatMul} latency and resource utilization.
Three key insights are drawn from these results.
First, since the \texttt{Rot} module is the main performance bottleneck, allocating additional hardware resources to the \texttt{CCadd} and \texttt{PCmul} modules has minimal impact on total latency.
Therefore, the \texttt{Rot} module is prioritized in resource allocation.
Second, the \texttt{CCadd} module uses significantly fewer hardware resources compared to other modules, allowing its $PC$ value to be large with negligible impact on overall utilization.
Finally, remaining resources are allocated to enable \texttt{PCmul} cores to process as many coefficients as possible.
Although $PC$ and $PI$ values differ across the four configurations, all are capable of processing 32 coefficients of a ciphertext polynomial in parallel.
From a resource perspective, increasing $PC$ is slightly more efficient than increasing $PI$, as it reduces the number of required \texttt{CCadd} cores.
Although not shown in Table \ref{tab:best_configuration}, when $PC$ exceeds 16, the \texttt{ct}$_\texttt{b}$ buffer is enlarged to support the increased bandwidth, reducing the URAM efficiency.
In such cases, increasing $PI$ becomes a better option.

Table \ref{tab:fpga_results} shows the resource utilization, as well as the latency, for each building homomorphic operator and the overall \texttt{MatMul} operator using the best parameter configuration.
\begin{table}[t]
	\centering
	\caption{FPGA Implementation Results (Target Freq. = 200MHz)}
	\label{tab:fpga_results}
	\begin{tabular}{l | c | c | c | c}
		\hline
        \multirow{2}{*}{Operator} & \multirow{2}{*}{DSP} & BRAM & URAM & Latency (ms) \\
        & & (36Kb) & (288Kb) & / Improvement \\
        \hline
        1 \texttt{CCadd} & 1 & 0 & 0 & 0.80 / 3.05$\times$ \\
        1 \texttt{PCmul} & 897 & 0 & 0 & 0.80 / 19.13$\times$ \\
        1 \texttt{Rot} & 5,123 & 1,536 & 313 & 31.35 / 6.81$\times$ \\
        \hline
        \texttt{MatMul} total & 6,920 & 1,536 & 659 & 2,150 / 13.86$\times$ \\
		\hline
	\end{tabular}
	\vspace{-1em}
\end{table}
To the best of the authors’ knowledge, this work is the first hardware accelerator for OMR.
Therefore, the latency results are compared with those obtained from a CPU-based implementation \cite{sophomr_github}, which are shown in Figs. \ref{fig:breakdown_a} and \ref{fig:breakdown_d}.
Note that in the default setting of \cite{sophomr_github}, the affine transform involves 2 \texttt{MatMul} operations, resulting in \texttt{MatMul} CPU time $\approx$ 29.8 seconds.
Our \texttt{MatMul} accelerator utilizes around 70\% of the FPGA resources and achieves a 13.86$\times$ speedup when scanning $2^{16}$ payloads, of which up to 50 are pertinent.

\section{Conclusion}
%
This paper presents an FPGA-based accelerator for the \texttt{MatMul} operation aimed at enhancing the practicality of OMR schemes.
We use an HLS technique to boost implementation productivity and improve accuracy in DSE.
In addition, we propose cost models to rapidly identify the optimal parameter configuration.
The implemented \texttt{MatMul} accelerator achieves a 13.86$\times$ speedup over the CPU implementation.
A limitation is the low performance of the \texttt{Rot} module.
Therefore, future work will focus on optimizing this component.

\section*{Acknowledgments}
%
This material is based upon work supported by the National Science Foundation under Grant No. 2347253.
The authors thank Dr. Young-kyu Choi for his valuable advice.

\bibliographystyle{IEEEtran}
\bibliography{references}

\end{document}